\documentclass{article}
\usepackage{graphicx}
\usepackage{amsfonts}

\begin{document}

\title{{\bf  Analytic Representation of Finite Quantum Systems}}

\author{S. Zhang and A. Vourdas\\
Department of Computing,\\
University of Bradford, \\
Bradford BD7 1DP, United Kingdom}
\maketitle

\begin{abstract}
A transform between functions in $\mathbb{R}$ and functions in $\mathbb{Z}_d$ is used to define the analogue of 
number and coherent states in the context of finite $d$-dimensional quantum systems. The coherent states are used to define 
an analytic representation in terms of theta functions. 
All states are represented by entire functions with growth of order $2$, which have exactly $d$ zeros in each cell.
The analytic function of a state is constructed from its zeros.
Results about the completeness of finite sets of coherent states within a cell, are derived.

\end{abstract}

\section{Introduction}

Quantum systems with finite Hilbert space have been studied originally by Weyl and Schwinger \cite{Weyl}
and later by many authors \cite{Auslander,Vourdas1}. A formalism analogous to the harmonic oscillator can be developed  
where the dual variables that we call `position' and `momentum' take values in 
$\mathbb{Z}_d$ (the integers modulo $d$). This area of research is interesting in its own right;
and has many applications in areas like quantum optics; quantum computing\cite{A000}; 
two-dimensional electron systems in magnetic fields and the magnetic 
translation group \cite{A100}; the quantum Hall effect \cite{A200}; quantum maps \cite{A300}; 
hydrodynamics \cite{A400}
mathematical physics; signal processing; etc.  
For a review see \cite{Vourdas2}.
 
In this paper we introduce a transform from functions in $\mathbb{R}$ to functions in $\mathbb{Z}_d$.
This is related to Zak transform from functions in $\mathbb{R}$ to functions on a circle 
\cite{Zak1,Zak2,Zak3}; but of course here we have functions in `discretized circle'.
This transform enables us to transfer some of the harmonic oscillator formalism into the context of finite systems.
For example, we define the analogues of number states and coherent states.
Coherent states in the context of finite systems have been previously considered in \cite{L1,coherent}.

Coherent states can be used to define analytic representations.
For example, ordinary coherent states of the harmonic oscillator can be used to define the Bargmann
analytic representation in the complex plane; $SU(1,1)$ coherent states can be used to define 
analytic representations in the unit disc (Lobachevsky geometry); 
$SU(2)$ coherent states can be used to define 
analytic representations in the extended complex plane (spherical geometry).
We use the coherent states in the context of finite systems to define analytic representations.
Similar analytic representations have been used in the context of quantum maps in \cite{L1}.
We show that the corresponding analytic functions obey doubly-periodic boundary conditions; 
and therefore it is sufficient to define them on a square cell $S$.
Each of these analytic functions has growth of order $2$ and has exactly $d$ zeros in $S$.

We use the analytic formalism to study the
completeness  of finite sets of coherent states in the cell $S$.
This discussion is the analogue in the present context, of the `theory of von Neumann lattice' 
for the harmonic oscillator \cite{V1,V2,V3,V4} ,
which is based on the theory of the density of zeros of analytic functions \cite{B}.

In section II, we review briefly the basic theory of finite systems and define some quantities for later use.
In section III, we introduce the transform between functions in $\mathbb{R}$ and functions in $\mathbb{Z}_d$.
Using this transform we define in section IV number states and coherent states in our context of finite systems, and study
their properties. 
In section V, we use the coherent states to define an analytic representation in terms of Theta functions.
We show that the order of the growth of these entire functions is $2$.
We also study displacements and the Heisenberg-Weyl group in this language. 
In section VI we study the zeros of the corresponding
analytic functions and use them to study the completeness of finite sets of coherent states within a cell.
In section VII we construct the analytic representation of a state from its zeros.
We conclude in section VIII with the discussion of our results.

\section{Finite quantum systems}
\subsection{Position and momentum states and Fourier transform}

We consider a quantum system with a $d$-dimensional Hilbert space ${\cal H}$. 
We use the notation $|s \rangle \rangle$ for the states in ${\cal H}$;
and we use the notation $|s \rangle $ for the states in the infinite dimensional Hilbert space $H$ associated with 
the harmonic oscillator.
Let  $|X;m \rangle \rangle$ be an orthonormal basis in ${\cal H}$,
 where $m$ belongs to $\mathbb{Z}_d$. 
We refer to them as `position states'. The $X$ in the notation is not a variable but it simply indicates
position states.

The finite Fourier transform is defined as:
\begin{eqnarray} \label{DFT1}
 F=d^{-1/2} \sum_{m,n} \omega(mn)|X;m\rangle\rangle \langle\langle X;n|;\;\;\;\;\;				
 \omega(\alpha)=\exp\left(i \frac{2\pi \alpha}{d}\right)
\end{eqnarray}
\begin{eqnarray} \label{DFT2}
 F F^{\dagger}=F^{\dagger} F={\bf 1};\;\;\;\;\;
 F^4={\bf 1}
\end{eqnarray}
Using the Fourier transform we define another orthonormal basis, 
the `momentum states', as: 
\begin{equation}
|P;m\rangle\rangle=F |X;m\rangle\rangle=d^{-1/2} \sum_{n=0}^{d-1} \omega(mn)|X;n\rangle\rangle
\end{equation}
We also define the `position and momentum operators' $x$ and $p$ as
\begin{eqnarray}
x=\sum _{n=0}^{d-1}n|X;n\rangle\rangle \langle\langle X;n|;\;\;\;\;\;\; 
p=\sum _{n=0}^{d-1}n|P;n\rangle\rangle \langle\langle P;n|
\end{eqnarray}
It is easily seen that
\begin{eqnarray}
F x F^{\dagger}=p;\;\;\;\;\;\;\;\;\;F p F^{\dagger}=-x
\end{eqnarray}

\subsection{Displacements and the Heisenberg-Weyl group}
The displacement operators are defined as:
\begin{equation}
Z=\exp \left (i \frac{2\pi}{d}x \right );\;\;\;\;\;
X=\exp \left (-i \frac{2\pi}{d}p \right )
\end{equation}
\begin{equation}
X^{d}=Z^{d}={\bf 1};\;\;\;\;\;\;
X^\beta Z^\alpha=Z^\alpha X^\beta \omega(-\alpha \beta)
\end{equation}
where $\alpha$,$\beta$ are integers in $\mathbb{Z}_d$. 
They perform displacements along the $P$ and $X$ axes in the $X-P$
phase-space.
Indeed we can show that:
\begin{equation}\label{movez}
Z^\alpha |P;m \rangle\rangle=|P; m+\alpha \rangle\rangle ;\;\;\;\;\;\;
Z^\alpha|X; m\rangle\rangle= \omega(\alpha m)|X; m\rangle\rangle
\end{equation}
\begin{equation}\label{movex}
X^\beta |P; m \rangle\rangle= \omega(-m\beta)|P; m \rangle\rangle ;\;\;\;\;\;
X^\beta|X; m\rangle\rangle=|X; m+\beta\rangle\rangle
\end{equation}
The $X-P$ phase-space is the toroidal lattice $\mathbb{Z}_d \times \mathbb{Z}_d$.

The general displacement operators are defined as:
\begin{equation} \label{displc}
D(\alpha, \beta)=Z^\alpha X^\beta \omega(-2^{-1}\alpha \beta);\;\;\;\;[D(\alpha, \beta)]^{\dagger}=D(-\alpha, -\beta)
\end{equation}
It is easy to see
\begin{eqnarray}  \label{dispx}
D(\alpha, \beta) |X;m \rangle\rangle &=& \omega (2^{-1}\alpha\beta + \alpha m) |X;m+\beta \rangle\rangle  \nonumber\\
D(\alpha, \beta) |P;m \rangle\rangle &=& \omega (-2^{-1}\alpha\beta - \beta m) |P;m+\alpha \rangle\rangle 
\end{eqnarray}

We next consider an arbitrary (normalized) state $|s \rangle\rangle $
\begin{eqnarray}
 |s\rangle\rangle = \sum_{m=0}^{d-1} s_m |X;m \rangle\rangle; \;\;\;\; \sum_{m=0}^{d-1} |s_m|^2 =1
\end{eqnarray}
and act with the displacement operators to get the states:
\begin{eqnarray}
 |s;\alpha, \beta \rangle\rangle \equiv D(\alpha, \beta) |s\rangle\rangle 
 = \sum_{m=0}^{d-1} s_m \omega(2^{-1}\alpha\beta + \alpha m) |X;m \rangle\rangle
\end{eqnarray}
Clearly $|s;0,0\rangle\rangle= |s\rangle\rangle$. Using Eq(\ref{movez}) and Eq(\ref{movex}) we easily show that
\begin{eqnarray}\label{genres}
 d^{-1} \sum_{\alpha,\beta=0}^{d-1} |s;\alpha, \beta \rangle\rangle \langle \langle s;\alpha, \beta | = \textbf{1}_d
\end{eqnarray}
This shows that the states $|s;\alpha, \beta \rangle\rangle $ (for a fixed `fiducial' state $|s\rangle\rangle $ and all $\alpha $, $\beta$ in
$\mathbb{Z}_d$ ) form an overcomplete basis of $d^2$ vectors in the d-dimensional Hilbert space ${\cal H}$. Eq(\ref{genres}) is the resolution of 
the identity. 

\subsection{General transformations}

In this section we expand an arbitrary operator  $\Omega$, in terms of displacement operators.
In order to do this we first define its
Weyl function as 
\begin{equation}
 \widetilde W_{\Omega}(\alpha,\beta) = \textrm{Tr} [\Omega D(\alpha,\beta)]
\end{equation}
The properties of the Weyl function and its relation to the Wigner function is discussed in \cite{Vourdas2}.
We can prove that
\begin{equation} \label{WelyDis}
 \Omega = d^{-1} \sum_{\alpha,\beta=0}^{d-1} \widetilde W_{\Omega}(-\alpha,-\beta) D(\alpha,\beta)
\end{equation}

\section{A transform between functions in  $\mathbb{R}$ and functions in  $\mathbb{Z}_d$}

In this section we introduce a 
map between states in the infinite dimensional harmonic oscillator 
Hilbert space $H$ and the $d$-dimensional Hilbert space ${\cal H}$. 
This map is a special case of the Zak transform. 
We consider a state $|\psi\rangle$
in $H$ with (normalized) wavefunction in the x-representation $\psi (x)=\langle x|\psi\rangle$.
The corresponding state $|\psi\rangle \rangle$ in ${\cal H}$ is defined through the map
\begin{eqnarray}
 \psi_m = \langle\langle X;m| \psi \rangle\rangle  = {\cal N}^{-1/2} \sum_{w=-\infty}^{\infty} 
          \psi \left[ x=\left( \frac{2\pi}{d} \right)^{1/2} \lambda (m+dw) \right];\;\;\;\;\;\;
          \psi_{m+d}=\psi_m 
  \label{mapx}
\end{eqnarray}
where $m\in \mathbb{Z}_d$. ${\cal N}$ is a normalization factor so that $\sum_{m=0}^{d-1} |\psi_m|^2 = 1$
which is given in appendix A. 
The Fourier transform (on the real line) of $\psi (x)$, is defined as:
\begin{eqnarray}
 \tilde\psi (p) &=& (2\pi)^{-1/2} \int_{-\infty}^{\infty} \psi (x) \exp(-ipx) dx. \label{Fho}
\end{eqnarray}
Using the map of Eq(\ref{mapx}) we define
\begin{eqnarray}
 \tilde\psi_m = {{\cal N}^\prime}^{-1/2} \sum_{w=-\infty}^{\infty} 
    \tilde\psi \left[ p=\left( \frac{2\pi}{d} \right)^{1/2} \frac{1}{\lambda} (m+dw) \right].  \label{mapp}
\end{eqnarray}
The tilde in $ \tilde\psi$ indicates that the Fourier transform of $\psi (x)$ has been transformed according to 
Eq.(\ref{Fho}). The normalization factor ${\cal N}^\prime$ is given in appendix A 
where it is shown that ${\cal N}^\prime=\lambda^2  {\cal N}$. 

We next prove that
\begin{eqnarray} \label{eigDFT}
\tilde\psi_m=  d^{-1/2} \sum_{n=0}^{d-1} \omega(-mn) \psi_n  = \langle\langle P;m| \psi \rangle\rangle 
\end{eqnarray}
This shows that $\tilde \psi_m $ is the finite Fourier transform of $\psi _n$, and therefore
the tilde also indicates the above finite Fourier transform.
So the tilde in the notation is used for two different Fourier transforms, but they are consistent to each other.

In order to prove Eq.(\ref{eigDFT}) we insert Eq.(\ref{mapx}) into Eq.(\ref{eigDFT})
and use the Poisson formula
\begin{eqnarray} \label{comb}
 \sum_{w=-\infty}^{\infty} \exp (i2\pi wx) = 
      \sum_{k=-\infty}^{\infty} \delta (x-k),
\end{eqnarray}
where the right hand side is the `comb delta function'; and also the relation
\begin{eqnarray} \label{modelta}
 \frac{1}{d} \sum_{m=0}^{d-1} \omega[m(k-\ell)]  = \delta(k,\ell);\;\;\;\;k,\ell\in \mathbb{Z}_d
\end{eqnarray}
where $\delta(k,k^\prime)$ is a Kronecker delta. 
These two relations are useful in many proofs in this paper.

The above map is not one-to-one (the Hilbert space $H$ is infinite dimensional while the Hilbert space 
${\cal H}$ is $d$-dimensional). Therefore, Eq.(\ref{mapx}) cannot be inverted. 
In appendix B, we use the full Zak transform and introduce a family of $d$-dimensional Hilbert spaces 
${\cal H}(\sigma_1,\sigma_2)$ with twisted boundary conditions.
We show that the Hilbert space $H$ is isomorphic to the direct
integral of all the ${\cal H}(\sigma_1,\sigma_2)$ (with $0\le\sigma_1<1$, $0\le\sigma_2<1$)
and then an inverse to the relation (\ref{mapx}) can be found. 
However the formalism of this paper is valid only for ${\cal H}\equiv {\cal H}(0,0)$
(which has periodic boundary conditions).

\section{Quantum states}

\subsection{Number eigenstates}

In the harmonic oscillator, number states are eigenstates of the Fourier operator $\exp (ia^\dagger a)$
where $a, a^\dagger $ are the usual annihilation and creation operators.
In this section we apply the transformation of Eq. (\ref{mapx}) with $\lambda =1$ and we show that the resulting states are
eigenstates of the Fourier operator of Eq.(\ref{DFT1}). 

We consider the harmonic oscillator number eigenstates $|N\rangle $ whose wavefunction is
\begin{eqnarray} 
 \chi (x,N)= \langle x|N \rangle &=& \left(\frac{1}{\pi^{1/2} 2^N N!}\right)^{1/2} \exp \left( -\frac{1}{2}x^2 \right) H_N(x),
\end{eqnarray}
It is known that
\begin{eqnarray}\label{tr}
 \tilde\chi (x, N) &=& i^N \chi (x,N).
\end{eqnarray}
Using the transforms of Eq(\ref{mapx}) and Eq(\ref{mapp}) with $\lambda =1$, we find:
\begin{eqnarray} \label{number}
 \chi_{m} (N)= \langle\langle X;m|N \rangle\rangle 
         &=& {{\cal N}_n(N)}^{-1/2}  \sum_{w=-\infty}^{\infty} \chi \left[ x=(m+dw)\left( \frac{2\pi}{d}\right)^{1/2} ,N\right], 
                  \label{eigDFx} \\
 \tilde\chi_{m}(N)= \langle\langle P;m|N \rangle\rangle  
  &=& {{\cal N}_n(N)}^{-1/2}\sum_{w=-\infty}^{\infty} \tilde\chi \left[ x=(m+dw)\left( \frac{2\pi}{d}\right)^{1/2} ,N \right], \label{eigDFp}
\end{eqnarray}
where ${\cal N}_n(N)$ is the normalization factor for number eigenstates, given by Eq(\ref{norm}) with $\psi$ replaced by $\chi$.
Eq.(\ref{tr}) implies that
\begin{eqnarray}
 \tilde\chi_{m} (N)&=& i^N \chi_{m} (N)\label{eigrlt}.
\end{eqnarray}
Using this in conjunction with Eq(\ref{eigDFT}) we prove that
\begin{eqnarray} \label{FN}
  d^{-1/2} \sum_{n=0}^{d-1} \omega(-mn) \chi _n(N) &=& i^N\chi_m(N);\;\;\rightarrow \;\; F|N\rangle\rangle = i^N |N\rangle\rangle.
\end{eqnarray}
Therefore the vectors $\chi_m(N)$ are eigenvectors of the Fourier matrix. 
They have been studied in the context of Signal Processing in \cite{eigen}
Of course, the Fourier matrix is finite and only $d$ of these eigenvectors are linearly independent.
Therefore the set of all states $|N\rangle\rangle$ is highly overcomplete.
In general the number states $|N\rangle\rangle$ are not orthogonal to each other.

The Fourier matrix has four eigenvalues $i^k$ ($0 \le k \le 3$); and
all the states $|N= 4M+k \rangle\rangle$ correspond to the same eigenvalue $i^k$.
As an example, we consider the case $d=6$ and using Eq.(\ref{eigDFx}) we calculate the six eigenvectors .
Results are presented in table I (we note that  $|5\rangle\rangle =-|1\rangle\rangle$).

\begin{center}
  \begin{tabular}{cccccc}
   \multicolumn{6}{c}{\bf Table I. Eigenvectors of the Fourier Operator F with $d=6$} \\ \hline \hline
     \hspace{4mm} $|0\rangle\rangle$ \hspace{7mm} & $|1\rangle\rangle$  \hspace{7mm} 
                       & $|2\rangle\rangle$ \hspace{7mm} & $|3\rangle\rangle$ \hspace{7mm} 
                       & $|4\rangle\rangle$ \hspace{7mm} & $|6\rangle\rangle$  \\ \hline
                             \hspace{4mm} 0.75971 \hspace{7mm} & 0 \hspace{7mm}
                            & -0.52546 \hspace{7mm} & 0 \hspace{7mm} & 0.37040 \hspace{7mm} & -0.31449 \\ 

                             \hspace{4mm} 0.45004 \hspace{7mm} & 0.65328 \hspace{7mm}
                            & 0.34071 \hspace{7mm} & -0.27059 \hspace{7mm} & -0.37823 \hspace{7mm} & 0.28578 \\

     \hspace{4mm} 0.09373 \hspace{7mm} & 0.27060 \hspace{7mm}
                            & 0.48131 \hspace{7mm} & 0.65328  \hspace{7mm} & 0.37471 \hspace{7mm} & -0.15803 \\

                             \hspace{4mm} 0.01365 \hspace{7mm} & 0 \hspace{7mm}
                            & 0.16851 \hspace{7mm} & 0 \hspace{7mm} & 0.54393 \hspace{7mm} &  0.82934 \\

                            \hspace{4mm} 0.09373 \hspace{7mm} & -0.27060 \hspace{7mm}
                            & 0.48131 \hspace{7mm} & -0.65328 \hspace{7mm} & 0.37471 \hspace{7mm} & -0.15803 \\

                             \hspace{4mm} 0.45004 \hspace{7mm} & -0.65328\hspace{7mm}
                            & 0.34071 \hspace{7mm} & 0.27059 \hspace{7mm} & -0.37823 \hspace{7mm} & 0.28578 \\ \hline
   \end{tabular}\\
\end{center}

\subsection{Coherent states}
We consider  the harmonic oscillator coherent states $|A \rangle$ whose wavefunction is
\begin{eqnarray}
 \psi(x,A)= \langle x|A \rangle  &=& \pi^{-1/4} \exp \left( -\frac{1}{2} x^2 + Ax - \frac{1}{2} A_R A \right), \label{csx}
\end{eqnarray}
where $A=A_R +iA_I$. Using the transformation of Eq(\ref{mapx}) we introduce coherent states $|A \rangle\rangle $
in the finite Hilbert space as:
\begin{eqnarray} \label{fcs}
 \psi_m(A)&=& \langle\langle X;m|A \rangle\rangle  = {{\cal N}_C(A)}^{-1/2}  \pi^{-1/4}
           \exp \left[ -\frac{\pi \lambda^2 m^2}{d} + A m \lambda \left( \frac{2\pi}{d} \right)^{1/2} - \frac{1}{2}A_R A \right ] \nonumber \\
   &&      \times \Theta_3 \left [ i\pi m \lambda^2 -i A \lambda \left( \frac{d \pi}{2} \right)^{1/2} ; id \lambda^2 \right ] \nonumber \\
   &=& {{\cal N}_C(A)}^{-1/2}  \pi^{-1/4} d^{-1/2}  \lambda^{-1}
           \exp \left( \frac{i}{2}A_IA \right ) 
           \cdot \Theta_3 \left [ \frac{\pi m}{d}-\frac{A}{\lambda} \left( \frac{\pi}{2d} \right)^{1/2}; \frac{i}{d \lambda^2} \right ].
\end{eqnarray}
where $\Theta _3$ are theta functions \cite{theta}, defined as
\begin{eqnarray}
 \Theta_3 (u;\tau) &=& \sum_{n=-\infty}^{\infty} \exp(i\pi\tau n^2+i2nu).
\end{eqnarray}
and the relation:
\begin{eqnarray}
 \Theta_3 (u;\tau) &=& (-i\tau)^{-1/2} \exp\left(\frac{u^2}{\pi i \tau}\right) 
                       \cdot \Theta_3 \left( \frac{u}{\tau};-\frac{1}{\tau} \right ),
\end{eqnarray}
has been used in Eq.(\ref{fcs}).
The normalization factor is:
\begin{eqnarray}
  {\cal N}_C(A) &=& \pi^{-1/2} \lambda^{-2} 
                  \cdot \left\{ \Theta_3\left[  \frac{A_R}{\lambda} (2\pi d)^{1/2}; \frac{2id}{\lambda^2} \right]
                              \Theta_3\left[  A_I i \lambda^{-1} \left( \frac{2 \pi}{d}\right)^{1/2};\frac{2i}{d\lambda^2} \right]
                   \right. \nonumber\\
                && + \left. \Theta_2\left[  \frac{A_R}{\lambda} (2\pi d)^{1/2}; \frac{2id}{\lambda^2} \right]
                            \Theta_2\left[  A_I i \lambda^{-1} \left( \frac{2 \pi}{d}\right)^{1/2};\frac{2i}{d \lambda^2} \right]
                    \right\}
\end{eqnarray}

The $\psi_m(A)$ obeys the relations:
\begin{eqnarray} \label{quasi}
 \psi_m \left[ A+(2\pi d)^{1/2} \lambda \right] &=& \psi_m(A) \exp \left[ i A_I \lambda \left( \frac{\pi d}{2}\right)^{1/2} \right];\nonumber \\
 \psi_m \left[ A+i \frac{(2\pi d)^{1/2}}{\lambda}\right]  &=& \psi_m(A) \exp \left[ -i \frac{A_R}{\lambda} \left( \frac{\pi d}{2}\right)^{1/2} \right]. 
\end{eqnarray}
The zeros of the Theta function $\Theta_3(u;\tau)$ are given by:
\begin{eqnarray} \label{thetazero}
 u = (2k+1)\frac{\pi}{2}+(2l+1)\frac{\pi\tau}{2},
\end{eqnarray}
where $k$,$l$ are integers. Therefore 
\begin{equation}
  \psi_m(A_{kl})= \langle\langle X;m|A _{kl}(m)\rangle\rangle =0;\;\;\;\
  A_{kl} (m)= \left( \frac{2\pi}{d} \right)^{1/2} \left[ \left( kd+\frac{d}{2}+m \right) \lambda +\frac{(2l+1)i}{2\lambda} \right].
\end{equation}
It is seen that the states $|A_{kl}(m)\rangle\rangle$ are orthogonal to the position states $|X;m\rangle\rangle$. 

The `vacuum state' $|0 \rangle\rangle$ is defined as
\begin{equation} \label{fcs7}
 \langle\langle X;m|0 \rangle\rangle = {{\cal N}_C(0)}^{-1/2} \pi^{-1/4} d^{-1/2}  
           \cdot \Theta_3 \left( \frac{\pi m}{d}; \frac{i}{d \lambda^2 } \right),
\end{equation}
where
\begin{eqnarray}
  {\cal N}_C(0) &=& \pi^{-1/2} \lambda^{-2} 
                  \cdot \left\{ \Theta_3\left[  0; \frac{2id}{\lambda^2} \right]
                              \Theta_3\left[  0;\frac{2i}{d\lambda^2} \right]
                     + \Theta_2\left[  0; \frac{2id}{\lambda^2} \right]
                            \Theta_2\left[  0;\frac{2i}{d \lambda^2} \right]
                    \right\}
\end{eqnarray}

The coherent states $|A \rangle\rangle $ satisfy the following resolution of the identity
\begin{eqnarray} \label{res}
 \lambda (2\pi d)^{-1/2}\int _S d^2A
  {\cal N}_C(A) |A \rangle\rangle \langle\langle A | = \textbf{1}_d ;\;\;
  \;\;\;S=\left[ a, a+(2\pi d)^{1/2}\lambda \right)_R \times \left[ b, b+\frac{(2\pi d)^{1/2}}{\lambda} \right)_I
\end{eqnarray}
We integrate here over the cell $S$.
The periodicity of Eq.(\ref{quasi}) implies that the cell can be shifted everywhere in the complex plane and this is 
indicated with the arbitrary real numbers $a$, $b$. The proof of Eq(\ref{res}) is based on the resolution of the identity
for ordinary (harmonic oscillator) coherent states, in conjunction with the map of Eq(\ref{mapx}).

The set of all coherent states in the cell $S$ is highly overcomplete.
Indeed using Eq.(\ref{genres}) we easily show another resolution of identity
which involves only $d^2$ coherent states in the cell $S$:
\begin{eqnarray} \label{res2}
 d^{-1}\sum_{\alpha, \beta =0}^{d-1} \left. \left\vert A+\left( \frac{2\pi}{d} \right)^{1/2}(\beta \lambda + \frac{\alpha}{\lambda} i) \right\rangle \right\rangle
   \left \langle \left \langle  A+\left( \frac{2\pi}{d} \right)^{1/2}(\beta\lambda + \frac{\alpha}{\lambda} i) \right\vert \right . = \textbf{1}_d
\end{eqnarray}

We calculate overlap of two coherent states
odd,
\begin{eqnarray}
 \langle\langle A_1|A_2 \rangle\rangle &=&  \pi^{-1/2} \lambda^{-2} {\cal N}_C(A_1)^{-1/2} {\cal N}_C(A_2)^{-1/2}
    \exp\left( -\frac{i}{2}A_{1I}A_1^\ast + \frac{i}{2}A_{2I}A_2 \right) \nonumber \\
   && \times \left \{ \Theta_3\left[ \frac{A_1^\ast+A_2}{\lambda} \left( \frac{\pi d}{2}\right)^{1/2}; \frac{2id}{\lambda^2}  \right]
      \Theta_3\left[ \frac{A_1^\ast-A_2}{\lambda} \left( \frac{\pi}{2d}\right)^{1/2}; \frac{2i}{d\lambda^2}  \right] \right. \nonumber \\
   && + \left. \Theta_2\left[ \frac{A_1^\ast+A_2}{\lambda} \left( \frac{\pi d}{2}\right)^{1/2}; \frac{2id}{\lambda^2}  \right]
      \Theta_2\left[ \frac{A_1^\ast-A_2}{\lambda} \left( \frac{\pi}{2d}\right)^{1/2}; \frac{2i}{d\lambda^2}  \right] \right\}.
\end{eqnarray}
and particularly find that when $d$ is an even number
\begin{eqnarray}
 \langle\langle A_1|A_2 \rangle\rangle &=&  \pi^{-1/2} \lambda^{-2} {\cal N}_C(A_1)^{-1/2} {\cal N}_C(A_2)^{-1/2}
    \exp\left( -\frac{i}{2}A_{1I}A_1^\ast + \frac{i}{2}A_{2I}A_2 \right) \nonumber \\
   && \times \Theta_3\left[ \frac{A_1^\ast+A_2}{\lambda} \left( \frac{\pi d}{8}\right)^{1/2}; \frac{id}{2\lambda^2}  \right]
      \Theta_3\left[ \frac{A_1^\ast-A_2}{\lambda} \left( \frac{\pi}{2d}\right)^{1/2}; \frac{2i}{d\lambda^2}  \right].
\end{eqnarray}
In the case, the first theta function in the above relation is zero when
\begin{eqnarray}
A_2-A_1^\ast=\left ( l+ \frac{1}{2} \right) (2\pi d)^{1/2} \lambda + i(2k+1)\left( \frac{2\pi}{d} \right)^{1/2} \lambda^{-1}
\end{eqnarray}
and the second is zero when
\begin{eqnarray}
A_2+A_1^\ast= \left(\frac{2 \pi}{d} \right)^{1/2} \lambda ( 2k+1 )
             + \left(\frac{\pi d}{2} \right)^{1/2} \lambda^{-1} (2l+1)i
\end{eqnarray}
The corresponding coherent states in these two cases are orthogonal to each other.

There is a relation between the coherent states in a finite Hilbert space studied in this section and the number states 
studied earlier:
\begin{equation}\label{nu}
 |A\rangle\rangle = \exp \left(-\frac{|A|^2}{2} \right) \sum_{N=0}^{\infty} \frac{A^N}{\sqrt{N!}} 
    \left[  \frac{{\cal N}_n(N)}{{\cal N}_C(A)} \right]^{1/2} |N\rangle\rangle.
\end{equation}
This is analogous to the relation between coherent states and number states in the infinite dimensional Hilbert space
for the harmonic oscillator. We have explained earlier that only $d$ of the number states appearing in the right hand 
side of Eq. (\ref{nu}) are independent.

Introducing the displacement operator defined in Eq(\ref{displc}), we can prove that
\begin{eqnarray}
 D(\alpha, \beta)|A\rangle \rangle = \left. \left\vert A+\left( \frac{2\pi}{d} \right)^{1/2}(\beta \lambda + \alpha \lambda^{-1} i) \right\rangle \right\rangle 
    \cdot \exp \left[ -iA_I \lambda \left( \frac{\pi}{2d}\right)^{1/2} \beta + iA_R \lambda^{-1} \left( \frac{\pi}{2d}\right)^{1/2} \alpha \right],
\end{eqnarray}
where both $\alpha$ and $\beta$ are integers.
We might be tempted to use the above equation as a definition for displacement operators 
with real values of $\alpha$, $\beta$. It can be shown that in this case $D$ depends on $A$;
and only for integer $\alpha$, $\beta$ the $D$ is independent of $A$.

\section{Analytic Representation}

\subsection{Quantum states}
Let $|f\rangle\rangle$ be an arbitrary(normalized) state
\begin{eqnarray}\label{st}
 |f\rangle\rangle = \sum_{m=0}^{d-1} f_m |X;m\rangle\rangle \;\;\;\;\; \sum_{m=0}^{d-1} |f_m|^2 =1.
\end{eqnarray}
We shall use the notation 
\begin{eqnarray}
 |f^\ast \rangle\rangle = \sum_{m=0}^{d-1} f_m^\ast |X;m\rangle\rangle \nonumber\\
 \langle\langle f|=\sum_{m=0}^{d-1} f_m^\ast \langle \langle X;m| \nonumber \\
 \langle\langle f^*| = \sum_{m=0}^{d-1} f_m \langle \langle X;m|
\end{eqnarray}

We define the analytic representation of $|f\rangle\rangle$, as:
\begin{eqnarray}\label{ana}
 f(z)  &\equiv & {{\cal N}_C(z)}^{1/2} d^{1/2} \lambda \exp \left(-\frac{i}{2} z_Iz \right) \langle\langle z^\ast |f \rangle\rangle \nonumber\\
 &=& \pi^{-1/4} \sum_{m=0}^{d-1} \Theta_3 \left [ \frac{\pi m}{d}-\frac{z}{\lambda}\left( \frac{\pi}{2d}\right)^{1/2}; \frac{i}{d\lambda^{2}} \right ] f_m  \label{analy}
\end{eqnarray}
where $|z\rangle\rangle$ is a coherent state. It is easy to see
\begin{eqnarray} \label{periodicity}
 f\left[ z+ (2\pi d)^{1/2} \lambda \right] = f(z); \;\;\;
 \;\; f\left[ z+ i (2\pi d)^{1/2} \lambda^{-1} \right]  = f(z)\exp \left[ \frac{\pi d}{\lambda^2} - i (2\pi d)^{1/2}z\lambda^{-1} \right].
\end{eqnarray}
The $f(z)$ is an entire function. 
If $M(R)$ is the maximum modulus of $f(z)$ for $|z|=R$,then
\begin{equation}
\rho =\lim _{R\to \infty }\sup \frac {\ln \ln M(R)}{\ln R}
\end{equation}
is the order of the growth of $f(z)$ \cite{B}.
It is easily seen that in our case the order of the growth is $\rho =2$.

Due to the periodicity our discussion below is limited to a single cell $S$ (defined in Eq.(\ref{res}) ).
The scalar product is given by 
\begin{eqnarray}
 \langle\langle f^\ast| g \rangle\rangle = (2\pi)^{-1/2} d^{-3/2} \lambda^{-1} \int_S d^2z \exp \left( - z_I^2 \right) f(z) g(z^\ast).
\end{eqnarray}

As special cases, we derive the analytic representation of the position states:  
\begin{eqnarray}
 |X;m\rangle\rangle \;\; &\rightarrow \;\; & \pi^{-1/4}  
     \cdot \Theta_3 \left [ \frac{\pi m}{d}-z \lambda^{-1} \left( \frac{\pi}{2d}\right)^{1/2}; \frac{i}{d \lambda^{2}} \right ] 
\end{eqnarray}
momentum states:
\begin{eqnarray}
 |P;m\rangle\rangle \;\; &\rightarrow \;\;& \lambda \pi^{-1/4} \exp \left(-\frac{1}{2}z^2\right)
     \cdot \Theta_3 \left [ \frac{\pi m}{d}- \lambda zi\left( \frac{\pi}{2d}\right)^{1/2}; \frac{i\lambda^2}{d} \right ] 
\end{eqnarray}
and the coherent states:
\begin{eqnarray}
  |A\rangle\rangle \;\; &\rightarrow& \nonumber\\
  f(z,A) &=& \pi^{-1/2} \lambda^{-1} d^{1/2} {\cal N}_C(A)^{-1/2} \exp\left( \frac{i}{2} A_I A\right) \nonumber\\
        && \times \left \{ \Theta_3\left[ \frac{z+A}{\lambda} \left( \frac{\pi d}{2}\right)^{1/2}; \frac{2id}{\lambda^2}  \right]
      \Theta_3\left[ \frac{z-A}{\lambda} \left( \frac{\pi}{2d}\right)^{1/2}; \frac{2i}{d\lambda^2}  \right] \right. \nonumber \\
   && + \left. \Theta_2\left[ \frac{z+A}{\lambda} \left( \frac{\pi d}{2}\right)^{1/2}; \frac{2id}{\lambda^2}  \right]
      \Theta_2\left[ \frac{z-A}{\lambda} \left(
  \frac{\pi}{2d}\right)^{1/2}; \frac{2i}{d\lambda^2}  \right]
  \right\}.
\end{eqnarray}
Again, when $d$ is even, it can be simplified as 
\begin{eqnarray}
  f(z,A) &=& \pi^{-1/2} \lambda^{-1} d^{1/2} {\cal N}_C(A)^{-1/2} \exp\left( \frac{i}{2} A_I A\right) \nonumber\\
             && \times \Theta_3\left[ \frac{z+A}{\lambda} \left( \frac{\pi d}{8}\right)^{1/2}; \frac{id}{2\lambda^2}  \right]
      \Theta_3\left[ \frac{z-A}{\lambda} \left( \frac{\pi}{2d}\right)^{1/2}; \frac{2i}{d\lambda^2}  \right];
\end{eqnarray}

\subsection{Displacements and the Heisenberg-Weyl group}
In this section we express the displacement operators $X$ and $Z$ in the context of analytic representations.
Eqs.(\ref{movex}),(\ref{movez}) are written as
\begin{eqnarray}
 Xf(z)=f \left[ z-\left( \frac{2\pi}{d} \right)^{1/2} \lambda \right];  \;\;\;\;\;\;
 Zf(z)=f\left[ z+i\left( \frac{2\pi}{d} \right)^{1/2} \lambda^{-1} \right ] \exp\left[iz \lambda^{-1}\left( \frac{2\pi}{d} \right)^{1/2}-\frac{\pi}{d\lambda^{2}}\right]. 
\end{eqnarray}
Therefore $X$ and $Z$ are given by:
\begin{eqnarray}
    X&=&\exp \left[ -\left( \frac{2\pi}{d} \right)^{1/2} \lambda \partial_z \right] \nonumber\\
   Z&=&\exp \left[ iz \lambda^{-1}\left( \frac{2\pi}{d} \right)^{1/2}-\frac{\pi}{d\lambda^{2}}\right] 
                        \exp \left[ i\left( \frac{2\pi}{d} \right)^{1/2} \lambda^{-1} \partial_z \right]     
\end{eqnarray}
and the general displacement operator is:
\begin{eqnarray}
  D(\alpha , \beta ) = \omega (-2^{-1/2}\alpha \beta ) \exp\left[i\alpha z \lambda^{-1}\left( \frac{2\pi}{d} \right)^{1/2}-\frac{\alpha^2 \pi}{d\lambda^{2}}\right] 
                       \exp \left[ (i\alpha\lambda^{-1}-\beta\lambda ) \left( \frac{2\pi}{d} \right)^{1/2}\partial_z \right] 
\end{eqnarray}
where $\alpha, \beta$ are integers in $\mathbb{Z}_d$.
Acting with this operator on the state $|f\rangle\rangle$ of Eq.(\ref{st}) represented by the analytic function $f(z)$
of eq(\ref{ana}), we get
\begin{eqnarray}
  D(\alpha , \beta ) f(z) &=& \pi^{-1/4} \exp \left[ i\alpha z \lambda^{-1} \left( \frac{2\pi}{d} \right)^{1/2}
              - \frac{\alpha^2 \pi}{d\lambda^2} - \frac{2^{1/2} \alpha \beta \pi i}{d} \right] \nonumber \\
            && \times \sum_{m=0}^{d-1} f_m \Theta_3 \left[ \frac{\pi m}{d} - \frac{z}{\lambda} \left( \frac{\pi}{2d} \right)^{1/2}
             -  \frac{\alpha \pi i}{d\lambda} + \frac{\beta \pi}{d}; \frac{i}{d\lambda^{2}} \right].  
\end{eqnarray}

\subsection{General transformations}

We have seen in Eq.(\ref{WelyDis}) that an arbitrary operator $\Omega$ can be expanded in terms of displacement operators
and using this we can express $\Omega$ as:
\begin{eqnarray}
  \Omega &=& d^{-1} \sum_{\alpha, \beta=0}^{d-1} \omega (-2^{-1/2}\alpha \beta ) \widetilde W_{\Omega}(-\alpha, -\beta)
               \exp\left[i\alpha z \lambda^{-1} \left( \frac{2\pi}{d} \right)^{1/2}-\frac{\alpha^2 \pi}{d\lambda^2}\right] \nonumber \\
         &&              \exp \left[ (i\alpha \lambda^{-1}-\beta \lambda ) \left( \frac{2\pi}{d} \right)^{1/2}\partial_z \right] 
\end{eqnarray}

Alternatively the operator $\Omega = \sum_{m,n} \Omega_{mn}|X;m\rangle\rangle \langle\langle X;n|$ 
can be represented with the kernel
\begin{eqnarray}
  \Omega(z,\zeta^*) &\equiv& {{\cal N}_C(\zeta)}^{1/2}{{\cal N}_C(z)}^{1/2} \lambda^2 
          \exp \left( -\frac{i}{2}z_I z + \frac{i}{2}\zeta_I \zeta^* \right)
                        \langle\langle z^* |\Omega| \zeta^* \rangle\rangle \\
  &=& \pi^{-1/2}d^{-1} \sum_{m,n=0}^{d-1} \Omega_{mn}
         \Theta_3\left[ \frac{\pi m}{d}-\frac{z}{\lambda}\left( \frac{\pi}{2d} \right)^{1/2}; \frac{i}{d\lambda^2} \right]
         \Theta_3\left[ \frac{\pi n}{d}-\frac{\zeta^*}{\lambda}\left( \frac{\pi}{2d} \right)^{1/2}; \frac{i}{d\lambda^2} \right]
\end{eqnarray}
and we easily prove that 
\begin{eqnarray}
  \Omega | f\rangle\rangle\rightarrow  (2\pi d)^{-1/2} \lambda^{-1} \int_S d^2 \zeta \exp \left( -{\zeta_I}^* \right)
                    \Omega (z,\zeta^*) f(\zeta)      
\end{eqnarray}
It is easily seen that
\begin{eqnarray}
  \Omega[z+(2\pi d)^{-1/2}\lambda \alpha,\zeta^*+(2\pi d)^{-1/2}\lambda \beta]&=& \Omega[z,\zeta^*]    \nonumber\\
  \Omega[z+i(2\pi d)^{-1/2}\lambda^{-1} \alpha,\zeta^*] &=& \Omega[z,\zeta^*]
         \exp \left[ \frac{\pi d}{\lambda^2}\alpha^2 -i(2\pi d)^{1/2}z\alpha \lambda^{-1} \right]  \nonumber\\
  \Omega[z,\zeta^*+i(2\pi d)^{-1/2}\lambda^{-1} \beta] &=& \Omega[z,\zeta^*]
         \exp \left[ \frac{\pi d}{\lambda^2}\beta^2 -i(2\pi d)^{1/2}\zeta^*\beta \lambda^{-1} \right]
\end{eqnarray}
where $\alpha$ and $\beta$ are integers. 

\section{Zeros of the analytic representation and their physical meaning}

If $z_0$ is a zero of the analytic representation $f(z)$, then Eq.(\ref{analy})
shows that the coherent state $|z_0\rangle\rangle$ is orthogonal to the state $|f\rangle\rangle$.

Using the periodicity of Eq(\ref{periodicity}) we easily prove that
\begin{equation}\label{1}
 \frac{1}{2\pi i} \oint_\Gamma \frac{f^\prime(z)}{f(z)} \; \textrm{d}z = d ,
\end{equation}
where $\Gamma$ is the boundary of the cell $S$. 
The above integral is in general equal to the number of zeros minus the number of poles of the function $f(z)$
inside $\Gamma$. Since our functions have no poles, we conclude that 
the analytic representation of any state has exactly $d$ 
zeros in the square $S$ (zeros will be counted with their multiplicities). The area of $S$ is $2\pi d$,
and therefore  there is an average of one zero per $2\pi$ area of the complex plane, in this analytic representation. 
As an example we show in Fig.1 the zeros of the coherent states $|0 \rangle \rangle$ and 
$|1+i \rangle \rangle$ for the case $d=4$.

\begin{figure}[p]
    \begin{center}
     \includegraphics[scale=0.8]{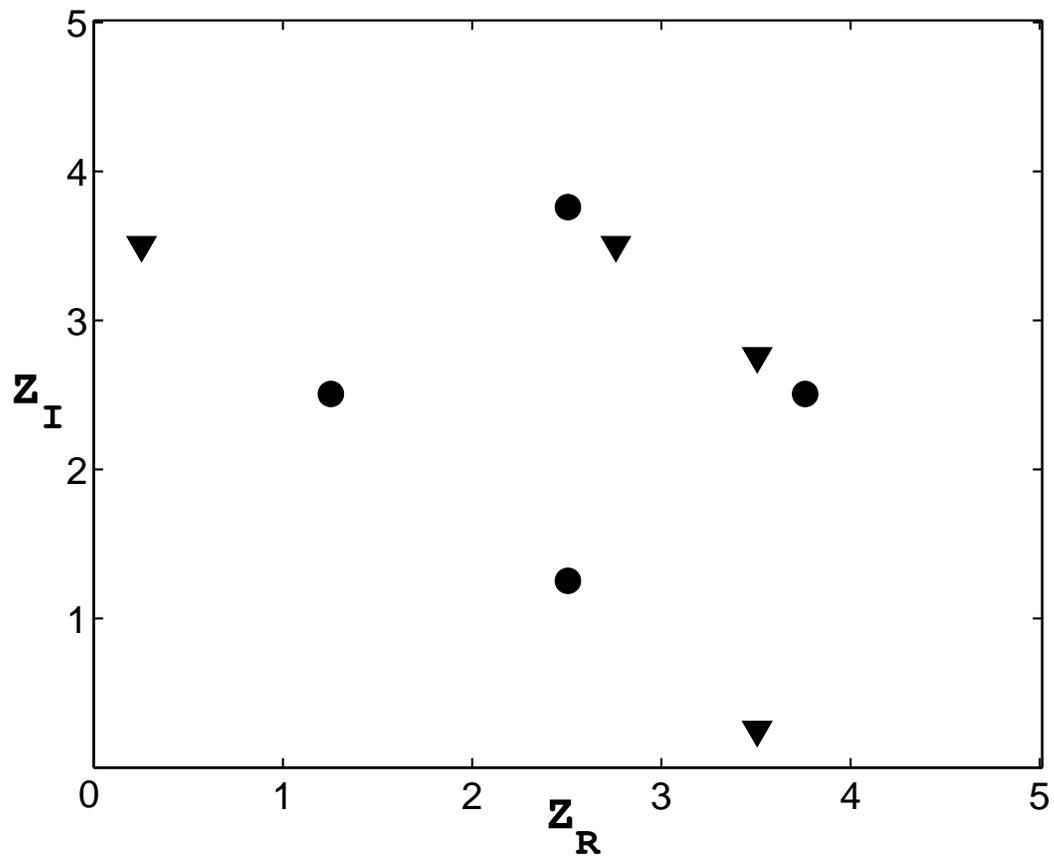}
     \caption{The zeros within a cell of the coherent states $|0 \rangle \rangle$ (circles) and 
$|1+i \rangle \rangle$ (triangles) for the case $d=4$.}
  
     \label{Fig1}
     
 \end{center}
\end{figure}

A direct consequence of this result is the fact that any set of $d+1$ coherent states in the cell $S$ is 
at least complete.
Indeed if it is not complete, then there exists some state which is orthogonal to all these coherent states.
But such a state would have $d+1$ zeros, which is not possible.
A set of $d+1$ states in a $d$-dimensional space which is at least complete is in fact overcomplete; 
in the sense that there exist a state which we can take out and be left with a complete set of $d$ states.
We note that if we take out an arbitrary state we might be left with an undercomplete set of $d$ states.

A set of $d-1$ coherent states is clearly undercomplete, because our Hilbert space is $d$-dimensional.

A set of $d$ distinct coherent states $\{|z_i\rangle\rangle; i=1,...,d \}$ will be complete or undercomplete depending on whether
it violates or satisfies the constraint
\begin{equation}\label{sumz}
 \sum _{i=1}^dz_i= \left( \frac{\pi}{2}\right)^{1/2} d^{3/2} (\lambda+i\lambda^{-1})
 +(2\pi d)^{1/2}(M\lambda+iN\lambda^{-1}).
\end{equation}
where $M,N$ are integers.
In order to prove this we use
the periodicity of Eq(\ref{periodicity}) to prove that
\begin{equation}\label{sumzero}
 \frac{1}{2\pi i} \oint_\Gamma \frac{z f^\prime(z)}{f(z)} \; \textrm{d}z = \left( \frac{\pi}{2}\right)^{1/2} d^{3/2} (\lambda+i\lambda^{-1})
 +(2\pi d)^{1/2}(M\lambda+iN \lambda^{-1}).
\end{equation}
The above integral is in general equal to the sum of zeros minus the sum of poles 
(with the multiplicities taken into account) of the function $f(z)$ inside $\Gamma$.
Since our functions have no poles, we conclude that the sum of zeros is equal to the right hand side of 
Eq(\ref{sumzero}). Eqs(\ref{1}),(\ref{sumzero}) have also been given in \cite{L1}.

If the $d$ coherent states considered violate Eq.(\ref{sumz}), then 
clearly they form a complete set because there exists no state which is orthogonal to all of them.
If however they do satisfy the constraint (\ref{sumz}), then there exists a state which is orthogonal to all of them.
To construct such a state we simply take the first $d-1$ coherent states $\{|z_1\rangle\rangle,...,|z_{d-1}\rangle\rangle\}$
(which form an undercomplete set because the space is $d$ -dimensional) and find a state $|g\rangle\rangle$
which is orthogonal to them. The corresponding analytic function $g(z)$ will have $d$ zeros which will be the
$z_1,...,z_{d-1}$ and an extra one which has to obey the constraint (\ref{sumz}) and therefore has to be $z_d$.
Therefore $|g\rangle\rangle$ will be orthogonal to $|z_d\rangle\rangle$ also, and consequently
the set of $\{|z_1\rangle\rangle,...,|z_d\rangle\rangle\}$ is undercomplete.

\section{Construction of the analytic representation of a state from its zeros}

We have proved in the last section that for an arbitrary state $|f\rangle\rangle$, the analytic representation $f(z)$
has $d$ zeros in the cell $S$ of Eq.(\ref{res}). 
In this section we assume that  the zeros $z_1,z_2,...z_d$ in the  cell $S$,
are given (subject to the constraint of Eq(\ref{sumz}))
and we will construct the function $f(z)$. 
We note that some of the zeros might be equal to each other.

We first consider the product 
\begin{equation}
  Q(z) = \prod_{j=1}^d  \Theta_3 \left[ (z-z_j+w) \left( \frac{\pi}{2d} \right)^{1/2} \lambda^{-1}; \frac{i}{\lambda^2} \right]
  ;\;\;\;\;\;w= \left( \frac{\pi d}{2} \right)^{1/2} (\lambda+\lambda^{-1}i)
\end{equation}
It is easily seen that $Q(z)$ has the given zeros.
The ratio $f(z)/Q(z)$ is entire function with no zeros and therefore it is the exponential of an entire function:
\begin{equation}
f(z)=Q(z)\exp (P(z))
\end{equation}
Taking into account the periodicity constraints of Eq.(\ref{periodicity}) we conclude that
\begin{eqnarray} \label{periodicity1}
 P\left[ z+ (2\pi d)^{1/2} \lambda \right]& = &P(z)+i2\pi K\nonumber\\ 
P\left[ z+ i (2\pi d)^{1/2} \lambda^{-1} \right] & = &P(z)+\frac {2\pi N}{\lambda ^2}+i2\pi \Lambda.
\end{eqnarray}
Here $N$ is the integer entering the constraint of Eq.(\ref{sumz});
and $K$,$\Lambda$ are arbitrary integers.
We have explained earlier that the growth of $f(z)$ is of order $2$.
The order of $Q(z)$ is $2$; therefore the $P(z)$ is a polynomial of maximum possible degree $2$.
Eq.(\ref{periodicity1}) shows that in fact $P(z)$ is 
\begin{eqnarray}
  P(z) = -\left( \frac{2\pi}{d} \right)^{1/2} N \lambda^{-1}z i +C.
\end{eqnarray} 
where $C$ is a constant. Therefore
\begin{eqnarray}\label{500}
  f(z) = C' \cdot \exp \left[ -\left( \frac{2\pi}{d} \right)^{1/2} N \lambda^{-1}z i \right]Q(z).
\end{eqnarray}
where the constant $C'$ is determined by the normalization condition.

\section{Discussion}

The harmonic oscillator formalism with phase space $\mathbb{R} \times \mathbb{R}$
has been studied extensively in the literature.
Equally interesting is quantum mechanics on a circle, with phase space $\mathbb{S} \times \mathbb{Z}$\cite{cir1,cir2};
and finite quantum systems, with phase space $ \mathbb{Z}_d \times \mathbb{Z}_d $.
Most of the results for physical systems on a circle or circular lattice (which is the case here), are 
intimately related to Theta functions; and well known mathematical results for Theta functions
can be used to derive interesting physical results for these systems.

In this paper we have introduced the transform of Eq.(\ref{mapx})
 between functions in $\mathbb{R}$ and functions in $\mathbb{Z}_d$.
The aim is to create a harmonic oscillator-like formalism in the context of finite systems.
 We have defined the analogue of 
number states for finite quantum systems in Eq.(\ref{number}); and of coherent states in Eq.(\ref{fcs}).
The properties of these states have been discussed.

Using the coherent states we have defined the 
analytic representation of Eq.(\ref{ana}) in terms of Theta functions.
In this language we have studied displacements and the 
Heisenberg-Weyl group; and also more general transformations.
Symplectic transformations are also important for these systems.
Especially in the case where $d$ is the power of a prime number, there are strong results (e.g., \cite{Vourdas2}).
Further work is needed in order to express these results in the language of analytic representations
used in this paper.

The analytic functions (\ref{ana}) have growth of order $2$ and they have exactly $d$ zeros
in each cell $S$. If the zeros are given we can construct the 
analytic representation of the state using Eq.(\ref{500}). 
Therefore we can describe the time evolution of a system
through the paths of the $d$ zeros of its analytic representation, in the cell $S$.

Based on the theory of zeros of analytic functions 
we have shown that a set of $d+1$ coherent states in the cell $S$ is overcomplete;
and a set of $d-1$ coherent states is undercomplete.
A set of $d$ coherent states in the cell $S$, is complete if the constraint of Eq.(\ref{sumz}) is violated;
and undercomplete if the constraint of Eq.(\ref{sumz}) is obeyed.
These results are analogous to the ` theory of von Neumann lattice'
in our context of finite quantum systems.

Our results use the powerful techniques associated to analytic 
representations in the context of finite systems.

\section{Appendix A}
The normalization factor appearing in Eq.(\ref{mapx}) is given by
\begin{eqnarray}
 {\cal N} = \sum_{m=0}^{d-1} \left \{ \sum_{w=-\infty}^{\infty} \psi^\ast \left[ x=\left( \frac{2\pi}{d} \right)^{1/2}\lambda(m+dw) \right]
           \right \} \left \{
            \sum_{w^\prime=-\infty}^{\infty} \psi \left[ x=\left( \frac{2\pi}{d} \right)^{1/2}\lambda(m+dw^\prime) \right] 
            \right \}\label{norm}
\end{eqnarray}
The normalization factor appearing in Eq.(\ref{mapp}) is given by
\begin{eqnarray}
 {\cal N}^\prime = \sum_{m=0}^{d-1} 
 \left \{ \sum_{w=-\infty}^{\infty} \tilde\psi^\ast \left[ p=\left( \frac{2\pi}{d} \right)^{1/2}\lambda^{-1}(m+dw) \right] \right \} 
 \left \{ \sum_{w^\prime=-\infty}^{\infty} \tilde\psi \left[ p=\left( \frac{2\pi}{d} \right)^{1/2}\lambda^{-1}(m+dw^\prime) \right] \right \} 
  \label{normp}
\end{eqnarray}
We insert Eq(\ref{Fho}) into Eq(\ref{normp}), and use Eqs(\ref{comb}), (\ref{modelta}) to prove that ${\cal N}^\prime=\lambda^{2}{\cal N}$.

\section{Appendix B}

In this appendix we use the full Zak transform to introduce a family of $d$-dimensional Hilbert space ${\cal H}(\sigma_1,\sigma_2)$
(with ${\cal H}\equiv {\cal H}(0,0)$). 
We generalize Eq.(\ref{mapx}) into
\begin{eqnarray}\label{genmap}
 \psi_m(\sigma_1,\sigma_2) =
        [{\cal N}(\sigma_1,\sigma_2)]^{-1/2}
                 \sum_{w=-\infty}^{\infty} \exp(-2\pi i \sigma_1 w) \psi \left[ \left( \frac{2\pi}{d} \right )^{1/2} \lambda
                       (m+\sigma_2 + dw) \right],
\end{eqnarray}
where ${\cal N}(\sigma_1,\sigma_2)$ is a normalization factor.
The Hilbert space ${\cal H}(\sigma_1,\sigma_2)$ is spanned by the states 
corresponding to $\psi_m(\sigma_1,\sigma_2)$. 
These spaces and the corresponding twisted boundary conditions of the wavefunctions, have been studied in \cite{L2}.
The Hilbert space $H$ is isomorphic to the direct
integral of all the ${\cal H}(\sigma_1,\sigma_2)$ (with $0\le\sigma_1<1$, $0\le\sigma_2<1$).
In this case Eq.(\ref{mapx}) can be inverted as follows:
\begin{equation}
 \psi \left[ x= \left( \frac{2\pi}{d}\right)^{1/2} \lambda (m+\sigma_2 + dw) \right] = \int_{0}^{1} [{\cal N}(\sigma_1,\sigma_2)]^{1/2}
         \psi_m(\sigma_1,\sigma_2)  \exp(2\pi i \sigma_1 w) \textrm{d}\sigma_1.
\end{equation}
The formalism of this paper is valid for the space ${\cal H}\equiv {\cal H}(0,0)$.

\end{document}